\documentclass[twocolumn,floatfix,prl,aps,showpacs,10pt]{revtex4}
\usepackage{graphicx,color,amsmath,amssymb}
\def\bc{\begin{center}}
\def\ec{\end{center}}
\def\beq{\begin{equation}}
\def\eeq{\end{equation}}
\def\bw{\begin{widetext}}
\def\ew{\end{widetext}}
\def\bea{\begin{eqnarray}}
\def\eea{\end{eqnarray}}

\renewcommand{\vec}[1]{{\textbf{\textit{#1}}}}

\usepackage[caption=false]{subfig}
\usepackage{amssymb,graphicx,dcolumn,bm,verbatim,epstopdf}

\begin{document}

\title{Quantum bubble defects in the lowest Landau level crystal}
\author{Alexander C. Archer and Jainendra K. Jain}
\affiliation{$^1$Department of Physics, 104 Davey Lab, The Pennsylvania State University, University Park, Pennsylvania 16802}

\date{\today}

\begin{abstract}
{
A longstanding puzzle for the lowest Landau level crystal phase has been an order of magnitude discrepancy between the theoretically calculated energy of the defects and the measured activation gap. We perform an extensive study of various kinds of defects in the correlated composite fermion crystal and find that the lowest energy defect is a six-fold symmetric  ``hyper-correlated bubble interstitial,"  in which an interstitial particle forms a strongly correlated bound state with a particle of the crystal. The energy of the bubble defect is a factor of $\sim$3 smaller than that of the lowest energy defect in a Hartree-Fock crystal. The anomalously low activation energies measured in transport experiments are thus a signature of the unusual quantum nature of the crystal and its defects.
}
\end{abstract}

\maketitle

When the extent of the quantum mechanical wave function of particles localized at the lattice sites of a crystal is comparable to the lattice constant, quantum mechanical effects can produce qualitatively new behavior.
That has inspired fascinating developments in the contexts of $^3$He and $^4$He solids \cite{Cross85,Leggett70}, as well as a longstanding interest in the crystal of electrons in the lowest Landau level (LLL) of a two-dimensional electron system (2DES) where the lattice constant is comparable to the magnetic length that governs the size of the electron wave packet.
(In contrast, for the electron crystal on the surface of liquid helium \cite{Grimes79} the quantum nature of electrons plays no measurable role.) The insulating phase observed at low fillings of the LLL \cite{Willett88,Goldman90,Jiang90,Jiang91,Paalanen92,Du96,Pan02,Santos92,Csathy04,Csathy05,Csathy07}) is believed to be a pinned crystal, and it is natural to ask in what ways the quantum nature of this crystal manifests in measurements. We show in this Letter that a striking feature of this crystal is the appearance of an unusual, non-classical defect with an extremely low energy.

\begin{figure}[t!]
{
\subfloat[]{\includegraphics[width=0.14\textwidth]{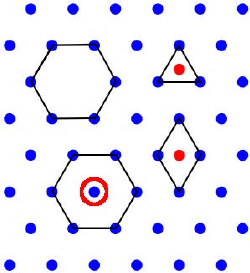}\label{fig:deffig}}

\subfloat[]{\includegraphics[width=0.24\textwidth]{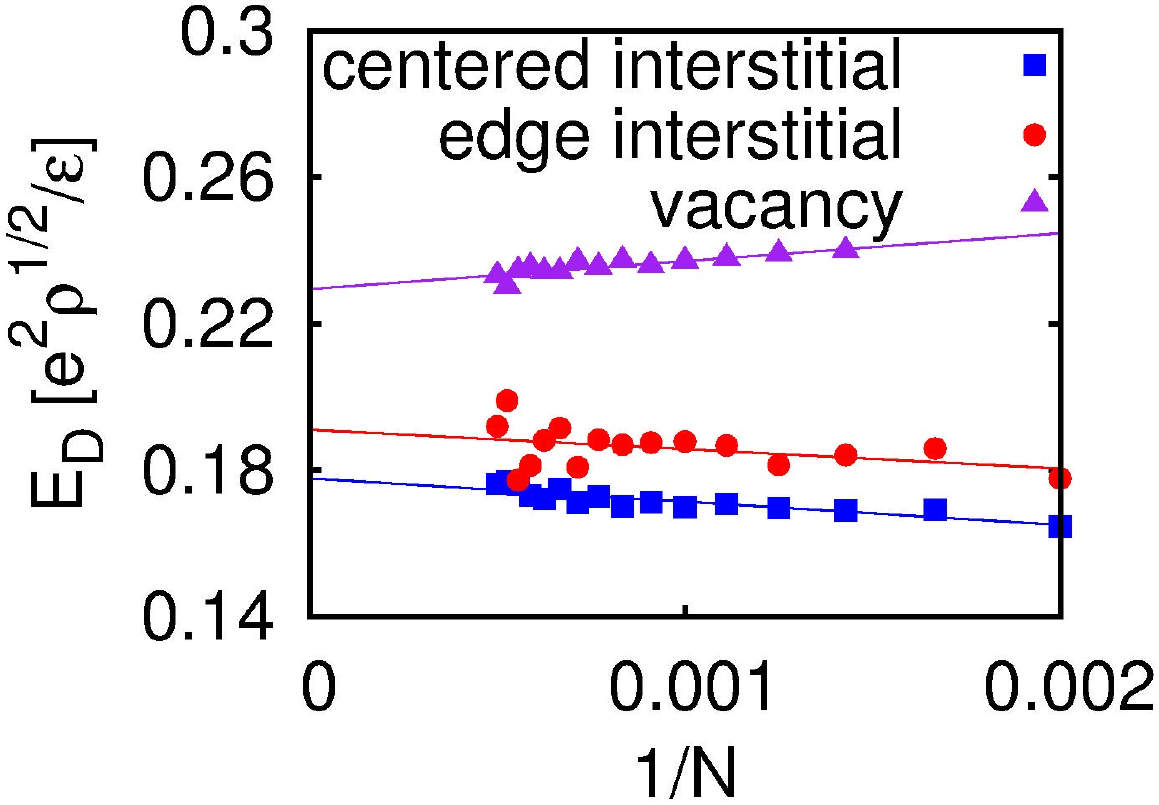}\label{fig:cl}}\hspace{-0.2cm}
\subfloat[]{ \includegraphics[width=0.23\textwidth]{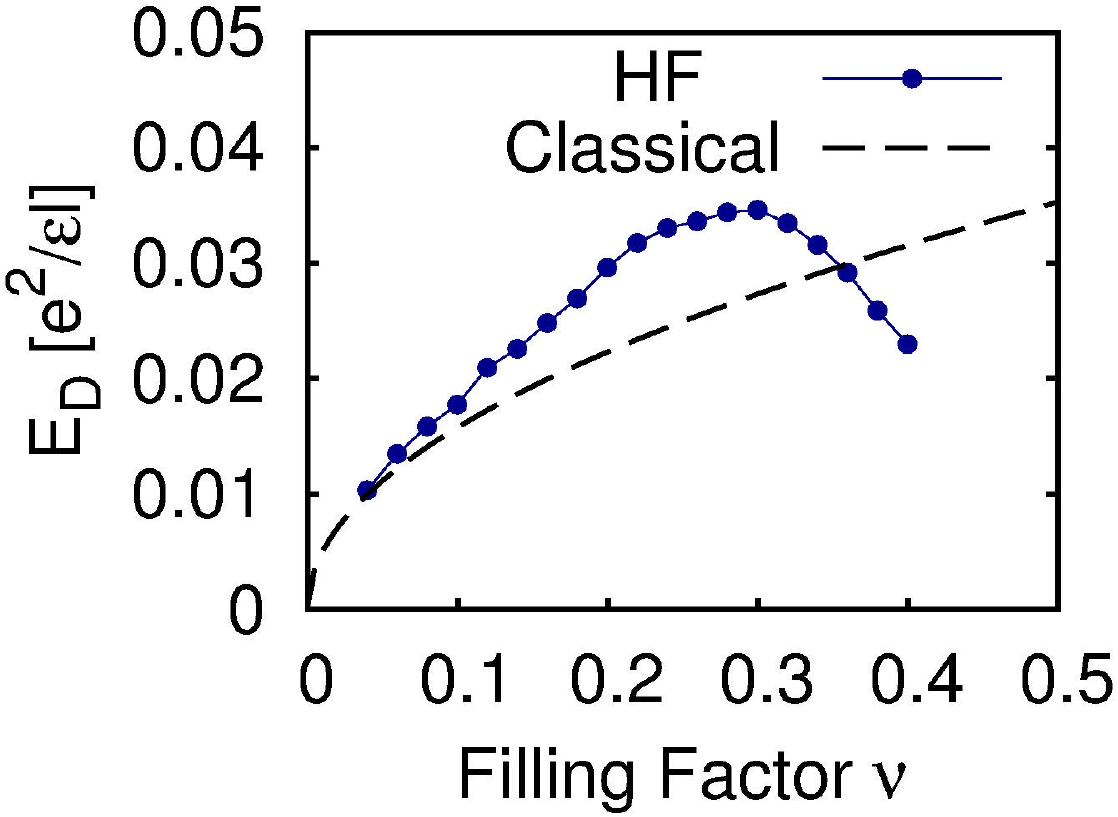}\label{fig:clmz}}
}
\caption{(a) A schematic representation of the defect configurations that we have considered. A vacancy is shown on top left. Going clockwise, the other three defects are centered, edge and bubble interstitials (with the added particle shown in red). The vacancy and the bubble interstitial have 6-fold symmetry, whereas the centered and edge interstitials have three and two fold symmetry, respectively. (b) The energies of various defects in a classical crystal, calculated using the radial relaxation procedure only, as a function of the system's size on the sphere. (c) The energy of the fully relaxed edge interstitial defect for the LLL Hartree Fock (HF) crystal (solid circles) and the classical crystal of point charges (dashed line) as a function of filling factor.}
\vspace{-0.4cm}
\label{fig:clhfeng}
\end{figure}

Surprisingly little theoretical work has been performed to investigate the lowest energy defects of the LLL crystal, even though it was was deduced experimentally more than two decades ago from the activated behavior of the longitudinal resistance, $\rho_{xx}\propto e^{\Delta/k_BT}$, in the insulating phase. The theoretical energies of various defects in a Hartree-Fock (HF) crystal, shown below, are found to be roughly an order of magnitude larger, thus indicating that the HF crystal is not a good description of the actual state. Indeed, theoretical work has shown that a crystal of composite fermions \cite{Yi98,Narevich01,Chang05,Archer11,Archer13} is superior to a HF crystal of electrons, and is also very close to the crystal formed in exact diagonalization studies \cite{Chang05}. Electrons thus take advantage of both the composite fermion (CF) and the crystal correlations to seek the lowest energy state: they bind fewer than the maximal number of vortices available to them and use the remaining degrees of freedom to form a crystal. We consider several point defects of the triangular CF crystal (CFC) and find the striking result that the lowest energy defect is not one of the standard point defects (i.e. a vacancy, an edge interstitial, or a centered interstitial) but rather a new kind of six-fold symmetric interstitial that we call a ``bubble interstitial," depicted in Fig.~\ref{fig:deffig}, in which the interstitial composite fermion forms a strongly correlated liquid bubble with one of the composite fermions forming the crystal; such a defect is not captured by elasticity theory because of its short range correlations. Furthermore, the CF bubble defect has a substantially lower energy than the lowest energy defect in the HF crystal. That leads us to suggest that the anomalously low energy of defects as deduced by transport experiments is a signature of the correlated quantum mechanical nature of the LLL crystal as well as its defects.

We define the defect energy $E_{\rm D}$ as \cite{Fisher79} 
\begin{equation}\label{eq:defeng}
 E_{\rm D}=\lim_{N\rightarrow\infty} \left(E^{(N)}_{\rm def}-E^{(N)}_0\right)
\end{equation}
where $E^{(N)}_{\rm def}$ is the energy of an $N$ particle crystal containing a defect and $E^{(N)}_0$ is the energy of a defect-free $N$ particle crystal. These energies include the interaction with the background and are evaluated at constant density (i.e. equal area). We find a smoother behavior as a function of $N$ by using the relation
\begin{equation}
 E_{\rm D}=\lim_{N\rightarrow\infty} \left(\frac{N}{N\pm1}E^{(N\pm 1)}_{\rm def}-E^{(N)}_0\right)
\end{equation}
because the lattice away from the defect is minimally changed between the crystals with and without defect. 

We perform our calculations in the spherical geometry \cite{Haldane83}. We form crystals on the surface of a sphere by placing wave packets at the Thomson minimum locations \cite{Thomson04,thomprob06} determined by minimizing the energy of $N$ charged point particles on the sphere. The sites are generally 6-fold coordinated, although the presence of some disclination defects is unavoidable in this geometry. The microscopic coordinates are denoted by $\vec{r}_j=(\theta_j,\phi_j)$ and the electron sites are denoted by $\vec{R}_l=(\gamma_l,\delta_l)$ in terms of the polar and the azimuthal angles. When describing the wave functions, it is convenient to use the spinor notation $(u_j,v_j)=(\cos(\theta_j/2)e^{i\phi_j/2},\sin(\theta_j/2)e^{-i\phi_j/2})$ and $(U_l,V_l)=(\cos(\gamma_l/2)e^{i\delta_l/2},\sin(\gamma_l/2)e^{-i\delta_l/2})$. We create a vacancy or an interstitial by removing a particle from a lattice site or by adding a particle to an interstitial site. Edge interstitial sites are given by $\vec{R}_{\rm edge}=(\vec{R}_1+\vec{R}_2)/|\vec{R}_1+\vec{R}_2|$ and centered interstitial sites are given by $\vec{R}_{\rm centered}=(\vec{R}_1+\vec{R}_2+\vec{R}_3)/|\vec{R}_1+\vec{R}_2+\vec{R}_3|$, where $\vec{R}_1$,$\vec{R}_2$, and $\vec{R}_3$ are any three neighboring lattice sites. A bubble interstitial \cite{Koulakov96,Fogler96} in a HF (CF) crystal consists of an electron (CF) pair localized at a Thomson lattice site.

Relaxing the crystal lattice around a defect is crucial for obtaining realistic defect energies. Past techniques for calculating the defect energy \cite{Fisher79,Price91,Cockayne91} rely on a periodic repetition of a large unit cell, and are not appropriate for use in the spherical geometry.
One must also take care not to relax the lattice by minimizing the energy of the entire system since this will simply heal the defect and produce a Thomson crystal with one more or fewer composite fermion. 
We have developed an efficient method for calculating the defect energy that proceeds along the following steps. We place the defect far from native disclinations on the sphere and first carry out `radial relaxation' which consists of a series of cycles during each of which we systematically allow successive sets of the defect's nearest neighbors to move either away from or toward the defect until the lowest energy is obtained. 
We find that for the system sizes considered in this article, the first relaxation cycle is the most important for the classical and HF crystals, reducing the defect energy by $\sim70\%$ for a vacancy and $\sim85\%$ for an interstitial; subsequent cycles produce a relatively small further reduction of $1\%-2\%$ for a vacancy and within $\sim5\%$ for an interstitial. Radial relaxation alone cannot be expected to produce the lowest possible defect energies, however. 
The complete relaxation procedure consists of relaxing the defect's first through fifth nearest neighbors using the conjugate gradient method \cite{Fisher79} (we found no further energy reduction by going to farther neighbors), relaxing the remainder of the lattice using radial relaxation, and obtaining the final defect energy by extrapolating to the thermodynamic limit. 
We have tested the effectiveness of this relaxation procedure by calculating defect energies for classical and LLL HF crystals (details below), where point charges interact through the Coulomb or the Maki-Zotos (MZ) interaction \cite{Maki83} (the MZ interaction $V_{\rm MZ}=\sqrt{\pi}I_0(r^2/8)\text{sech}(r^2/8)/4$, where $I_0$ is the modified Bessel function, gives the Coulomb interaction energy between two Gaussian wave packets in the LLL) for system containing up to $N=2000$ particles.   For a fixed filling factor in the spherical geometry, the density of a system has a slight $N$ dependence; the effect of this variation can be eliminated by making the  ``density correction" to total energy by multiplying it by a factor of $\sqrt{2Q\nu/N}$, which improves convergence of our results to the thermodynamic limit; we make this correction in all results.  We plot the defect energies for classical systems calculated using radial relaxation in Fig.~\ref{fig:cl} and the fully relaxed classical and HF edge interstitial defect energy in Fig.~\ref{fig:clmz}. The energies of a classical crystal in zero magnetic field are expressed in units of $e^2\sqrt{\rho}/\epsilon$ and those for the LLL crystal in units of $e^2/\epsilon\ell$, where $\ell=\sqrt{\hbar c/eB}$ is the magnetic length. We find that the classical defect energies calculated using this method are consistent with those found in previous studies \cite{Fisher79,Price91,Cockayne91}, producing centered and edge interstitial defect energies of $0.125$, and 0.19 for the vacancy defect. Comparison with Fig.~\ref{fig:clhfeng}b indicates that radial relaxation alone is not satisfactory, especially for the interstitials. 
The lowest energy defects for both classical and HF systems are interstitials, with there being only a small difference between edge or centered interstitial. We note that the HF defect has significantly higher energy than the corresponding classical defect for the filling factor range of interest (Fig.~\ref{fig:clhfeng}c).

\begin{figure}[t!]
{

\subfloat[$2p=0,q=0$]{\includegraphics[width=0.13\textwidth]{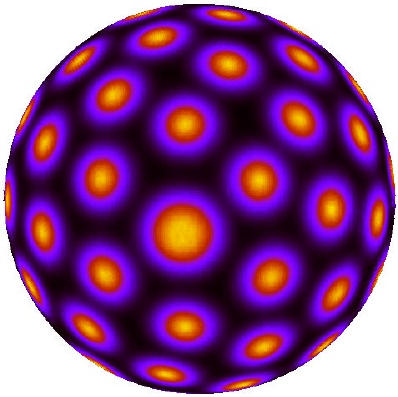}\label{fig:bd1}}
\hspace{0.3cm}\subfloat[$2p=2,q=0$]{\includegraphics[width=0.13\textwidth]{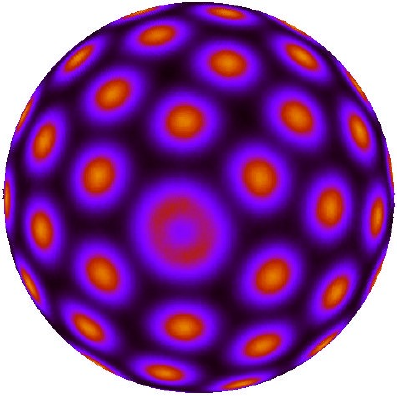}\label{fig:bd2}}
\hspace{0.2cm}\subfloat[$2p=2$, $q=1$]{ \includegraphics[width=0.18\textwidth]{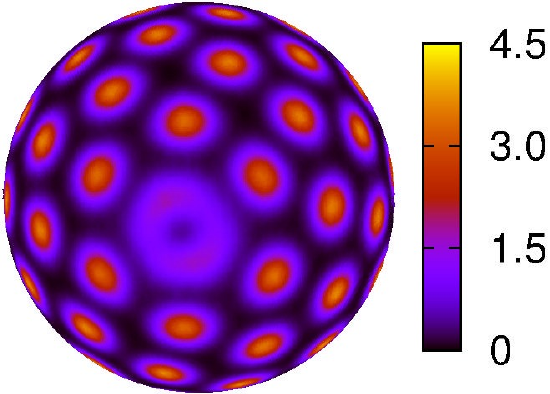}\label{fig:bd3}}

}
\caption{Density plot of a bubble interstitial in (a) a HF crystal, and (b) a CF crystal. Panel (c) shows a hyper-correlated bubble in a CF crystal. The plots are for a system of $N=64$ particles at $\nu=0.21$. The number of vortices attached to every electron, $2p$, and the number of additional vortices attached to the bubble interstitial, $q$, are shown. The densities are shown in units of the average density.}
\vspace{-0.6cm}
\label{fig:density}
\end{figure}

We now proceed to describe the crystal and defect wave functions for composite fermions. To construct the wave function for the defect-free CFC of composite fermions carrying $2p$ vortices, denoted $^{2p}$CFC, we place wave packets of electrons, $\phi_{\vec{R}_l}^{2Q^*}\!\!(\vec{r}_m)=(\widetilde{U}_lu_m+\widetilde{V_l}v_m)^{2Q^*}$, at the Thomson minimum locations \cite{thomprob06}, where $2Q^*$ is the effective monopole strength; anti-symmetrize the product; and then composite-fermionize by attaching $2p$ vortices to each particle \cite{Jain89,Jain07}. The final wave function is given by:
\begin{equation}\label{eq:cfcwf}
 \displaystyle \Phi_{\scriptstyle 2Q,\{\vec{R}\}}^{2p}\hspace{-0mm}=\prod_{j<k}^N(u_jv_k-v_ju_k)^{2p}\;A\!\left(
 \prod_{1\leq j \leq N}
 \phi_{\vec{R}_j}^{2Q^*}(\vec{r}_j)\right)
\end{equation}
where $N$ is the number of particles, the monopole strength is $2Q=2Q^*+2p(N-1)$, $\{\vec{R}\}\equiv \{\vec{R}_1,\vec{R}_2,\hdots,\vec{R}_N\}$ denotes the Thomson lattice sites, and $A$ is the anti-symmetrization operator. 

The wave function for an unrelaxed vacancy
is identical to that of the $N$ particle CFC except that a single lattice site is left unoccupied. 
For an unrelaxed interstitial, we insert an additional CF at $\vec{R}_I$. 
Following Zheng and Fertig \cite{Zheng94}, we also consider a `hyper-correlated' interstitial in which we attach $q$ extra vortices to the interstitial defect to build in additional repulsive correlations with the surrounding lattice. 
The resulting interstitial wave function is given by:
\begin{eqnarray}
 \displaystyle \Phi_{2Q,\{\vec{R}\},\vec{R}_I}^{2p,q} &&= \prod_{j<k}^{N+1}(u_jv_k-v_ju_k)^{2p} \nonumber \\
 &&A\!\left((J_1)^q\phi_{\vec{R}_{\rm I}}^{2Q'}\!(\vec{r}_1)
 \prod_{1\leq j \leq N}
  \phi_{\vec{R}_{j}}^{2Q''}\!(\vec{r}_{j+1})\right)
\end{eqnarray}
where $\displaystyle J_j=\prod_{k\neq j}(u_jv_k-v_ju_k)$, $2Q'=2Q^*-q(N-1)$, $2Q''=2Q^*-q$, and for a given $q$ we consider only those $2Q^*$ for which $2Q'>0$.

We also consider the bubble interstitial, which consists of two particles with relative angular momentum of $1$ localized at a Thomson lattice site. Similarly to hyper-correlated interstitials, one can create hyper-correlated bubble interstitials through the attachment of additional vortices, with its wave function given by:
\begin{eqnarray}\label{eq:bubdef}
\Phi_{2Q,\{\vec{R}\},\vec{R}_B}^{2p,q}&&=\prod_{j<k}^{N+1}(u_jv_k-v_ju_k)^{2p}\;A [(J_1)^q(J_2)^q \nonumber \\
 &&(u_1v_2-v_1u_2)\phi_{\vec{R}_{\rm B}}^{2Q'}\!(\vec{r}_1)\phi_{\vec{R}_{\rm B}}^{2Q'}(\vec{r}_2) \nonumber \\
&&  \prod_{\substack{1\leq j \leq N-1 \\ \vec{R}_j\in\{\vec{R}\}\setminus\ \{\vec{R}_B\}}}
  \phi_{\vec{R}_j}^{2Q''}\!(\vec{r}_{j+2})]\;
\end{eqnarray}
where $\vec{R}_B\in\{\vec{R}\}$ is the location of the defect pair, $2Q'=2Q^*-q(N-1)-q-1$, and $2Q''=2Q^*-2q$.

\begin{figure}[t]

\subfloat{\includegraphics[width=0.4\textwidth]{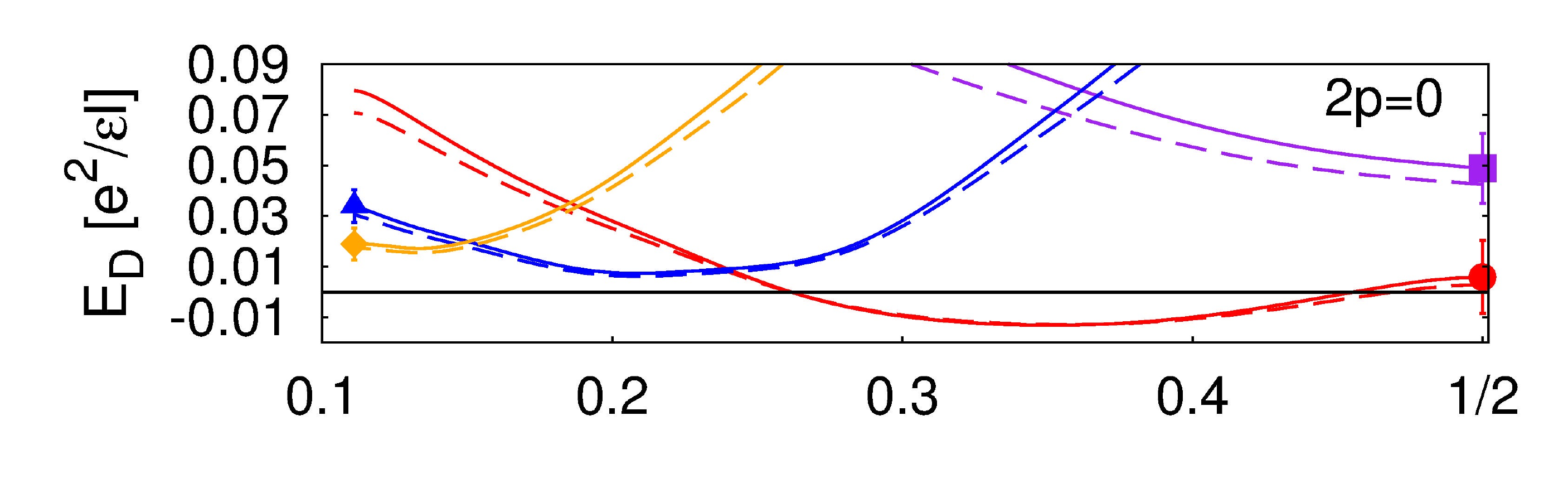}}
\vspace{-0.35cm}
\subfloat{\includegraphics[width=0.4\textwidth]{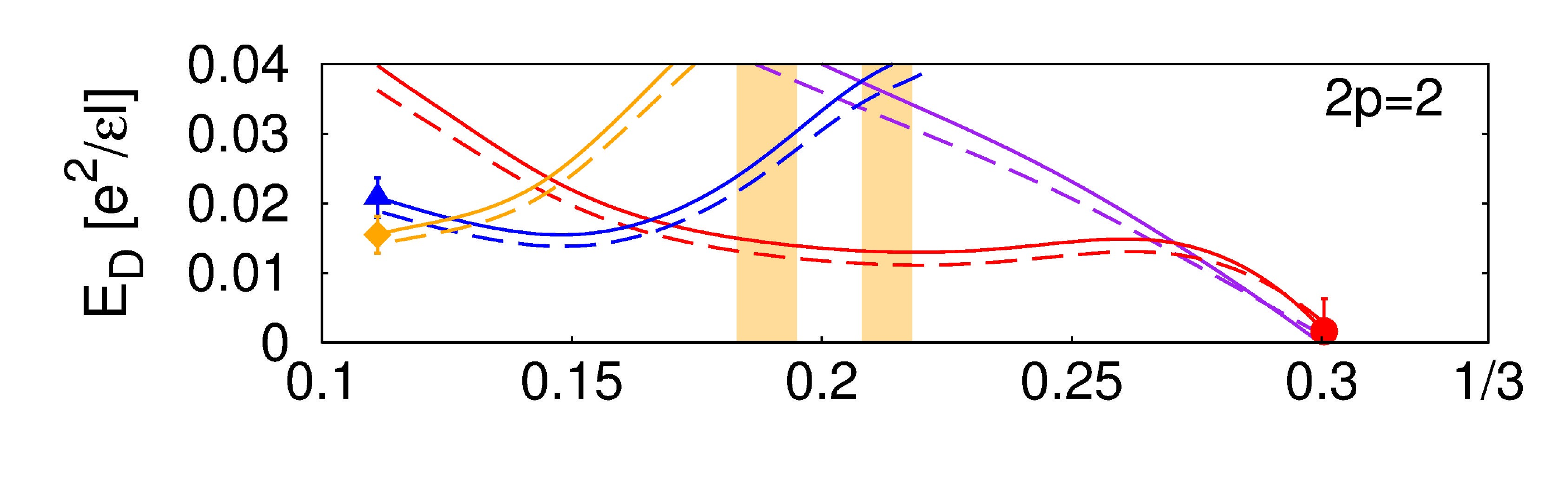}}
\vspace{-0.35cm}
\subfloat{\includegraphics[width=0.4\textwidth]{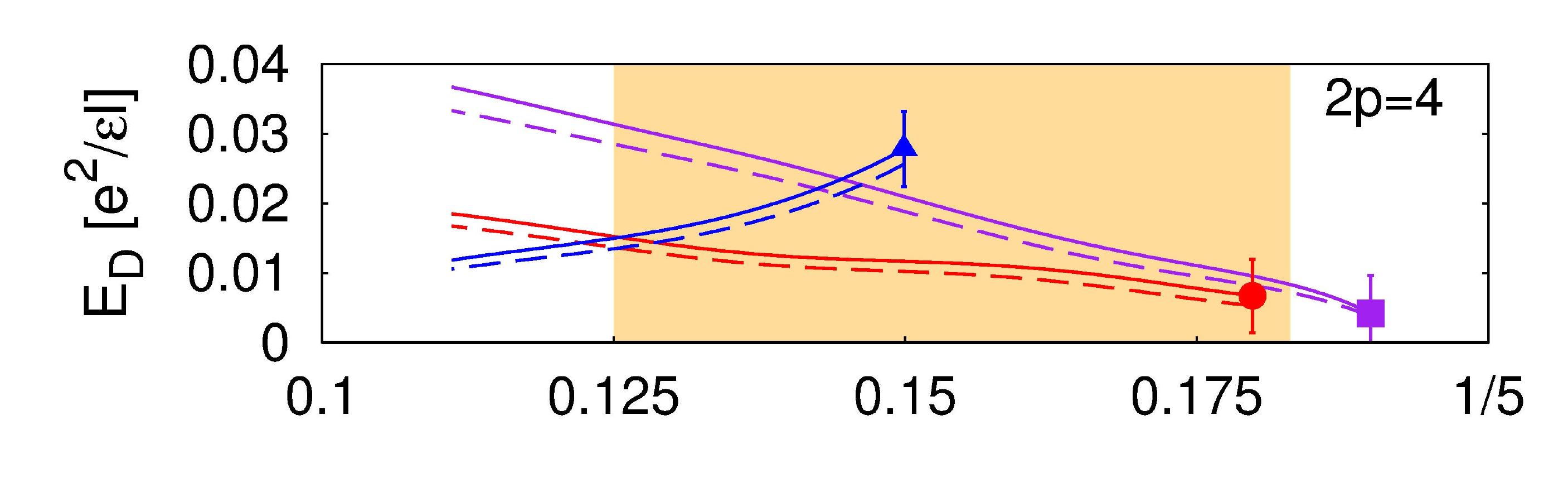}}
\vspace{-0.35cm}
\subfloat{\includegraphics[width=0.4\textwidth]{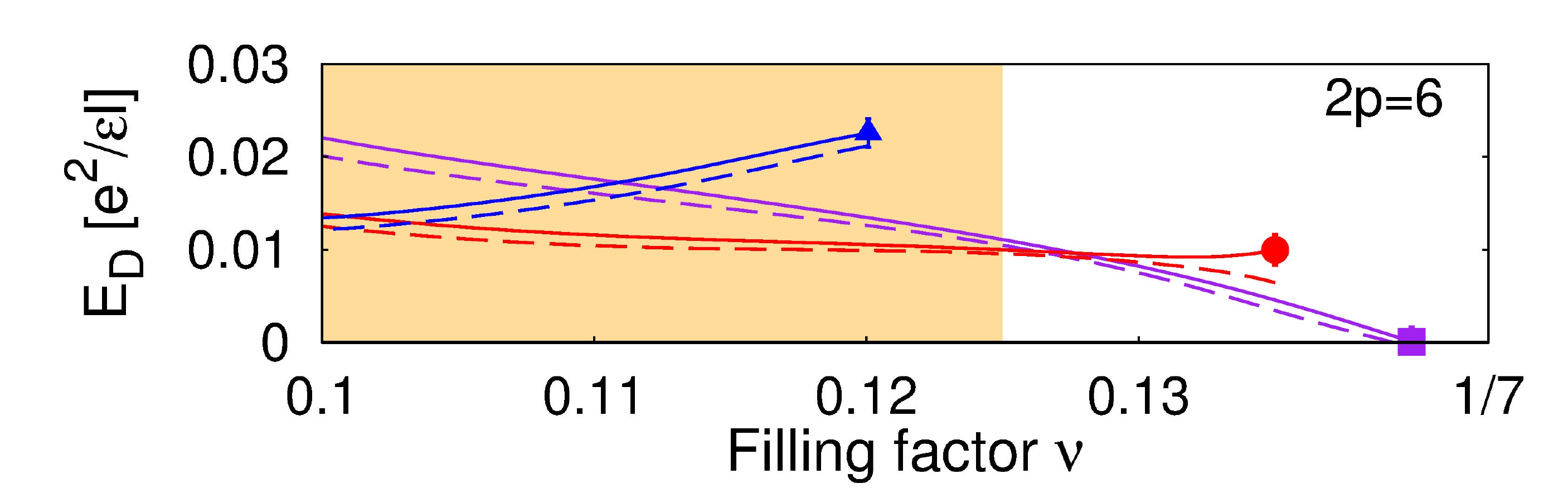}}

\caption{The bubble interstitial energy as a function of the filling factor for crystals of composite fermions with vorticity $2p=0$, 2, 4 and 6 (from top to bottom panel). Purple squares, red circles, blue triangles, and orange diamonds correspond to $q=0,1,2,$ and $3$, respectively, where $q$ denotes the number of additional vortices attached to the bubble, as discussed in the text. The HF bubble corresponds to $2p=0$ and $q=0$. The solid lines are for a system with zero thickness, whereas the dashed line are for a quantum well width of 40 nm with a density of $1.0\times 10^{11}$ cm$^{-2}$. The shaded regions correspond to the filling factor regions where the $^{2p}$CFC under consideration is the lowest energy crystal, according to the phase diagram derived by Archer, Park and Jain \cite{Archer13}. The lowest energies in the shaded regions are obtained for hyper-correlated bubbles with $q\geq 1$. The results are for a system with $N=65$ particles; error bars on each curve show the Monte Carlo statistical uncertainty in the energy.}
\vspace{-0.5cm}
\label{fig:defeng_all}
\end{figure}

We have performed an exhaustive evaluation of energies for the vacancy, interstitial, and bubble interstitial defects in $N=32$ and $64$ particle systems, for $2p=0,2,4,6$, and $q=0,1,2$, over a wide range of filling factors using standard Markov-Chain Monte Carlo techniques. To relax the CFC lattice surrounding the defect, we find it sufficient to perform a single cycle of radial relaxation; subsequent cycles only have a small effect. Remarkably, even radial relaxation causes only a very slight reduction in the defect energy for the $^{2p}$CFC, reducing the energies of vacancy and interstitial defects by no more than $20\%$ and producing essentially no change at all for the hyper-correlated bubble defect.  (In contrast, the energy of the classical centered interstitial goes from $\sim$1.48 to $\sim$0.18 after radial relaxation, and to 0.125 after full relaxation, with all energies in units of $e^2\rho^{1/2}/\epsilon$.) This shows that the correlations introduced by composite-fermionization automatically relax the surrounding lattice to a great degree. This also indicates that further relaxation would have a negligible impact on the CFC defect energies. We have also found that the defect energy has negligible dependence on its location on the sphere. In Fig.~\ref{fig:density} we show density plots for certain typical relaxed bubble defects.

The four panels of Fig.~\ref{fig:defeng_all} show the energies of various defects for $2p=0, 2, 4$, and 6. The physically relevant region for each CFC is highlighted in yellow in each panel, according to the phase diagram of $^{2p}$CFCs evaluated previously \cite{Archer13}.  In all cases we find, surprisingly, that the hyper-correlated $q=1$ bubble interstitial has the lowest energy. (This is surprising because the bubble interstitial has a rather high energy in the HF crystal.) Furthermore, the energy of the hyper-correlated bubble interstitial is a factor of three lower than the lowest energy HF defect near $\nu=1/5$. The energy of bubble interstitial in $N=64$ particle crystals is shown in Fig.~\ref{fig:defeng_all} as a function of the filling factor. (The other defects in the CFC have a higher energy, and will be described elsewhere.) We have accounted for the finite thickness of the 2DES by assuming an infinite square quantum well with cosine wave function, which produced an effective microscopic interaction given in Shi {\emph et al.} \cite{Shi08}; the finite thickness results are indicated by the dashed lines in Fig.~\ref{fig:defeng_all}. We have compared the bubble interstitial defect energies in $N=32$ and $64$ particles systems and found that the difference is smaller than the Monte Carlo statistical uncertainty, which leads us to believe that our results shown in Fig.~\ref{fig:defeng_all} are a good representation of the bubble interstitial defect energy in the thermodynamic limit. 

Experiments have measured the activation energy for the crystal surrounding the $\nu=1/5$ fractional quantum Hall (FQH) state \cite{Willett88,Goldman90,Jiang90,Jiang91,Paalanen92,Du96}. They find substantial filling factor dependence of the activation energy close to 1/5, but for filling factors sufficiently far below 1/5 this energy becomes approximately constant, and is given by 0.004 $e^2/\epsilon \ell$ in the highest mobility samples~\cite{Jiang91}. This is significantly lower than the energy of the lowest energy HF defect at $\nu=0.18$, which is 0.028 $e^2/\epsilon \ell$ (Fig.~\ref{fig:clmz}). It is in better agreement with the energy of the $q=1$ hyper-correlated bubble defect, which has an energy of approximately 0.010 $e^2/\epsilon \ell$ for the $^2$CF crystal surrounding the 1/5 state (Fig.~\ref{fig:defeng_all}).

A vanishing defect energy signals an instability of the crystal phase. Two kinds of instability may be seen in Fig.~\ref{fig:defeng_all}. The top panel illustrates the first type where we see that the energy of a $q=1$ hyper-correlated bubble in a $^0$CFC is negative for $\nu>0.3$; this indicates that the HF state is unstable to the formation of composite fermions. The bottom three panels of Fig.~\ref{fig:defeng_all} illustrate another type of instability, where the energy of $q=0$ bubble defect becomes negative as $\nu^*\rightarrow1$ ($\nu\rightarrow1/(2p+1)$); this signals the transition of the $^{2p}$CFC into an FQH state.

We note that our theory does not capture the rapid decay in the activation energy upon approach to $\nu=1/5$, as observed in experiments \cite{Jiang91}. Such behavior is indeed not expected for a first order transition from a crystal to liquid state. We attribute this behavior to the presence of disorder, which broadens the transition region while also converting it into a continuous one; this view is consistent with the fact that the transition region is sharpest in the highest mobility sample \cite{Jiang91}. Narevich, Murthy, and Fertig \cite{Narevich01} have treated the CFC in the vicinity of 1/5 with an effective Hamiltonian theory. They find that while $^4$CFC produces low defect energies with strong $\nu$ dependence as the system merges into the 1/5 FQH liquid (see the preceding paragraph), the defects in the $^2$CFC, which was found in Ref.~\onlinecite{Archer13} to be the relevant crystal in this filling factor range, are of much higher energy (greater than 0.1 $e^2/\epsilon \ell$ in the vicinity of $\nu=1/5$). 

We have neglected a number of features that can have a quantitative impact on the calculated defect energy. We have not accounted for Landau level mixing and disorder, which are expected to lead to a lowering of the gaps \cite{Cha94,Cha94b}.  
This suggests that improving sample quality should lead to yet higher experimental activation energies, a trend that is apparent in the experiments of Jiang \emph{et al.} \cite{Jiang90,Jiang91} and Du \emph{et al.} \cite{Du96}. While these effects are beyond the scope of this paper, it is worth noting that the discrepancy between the measured and theoretical defect energies is comparable to that between measured and theoretical gaps of the prominent FQH states, which is often attributed to a combination of Landau level mixing and disorder \cite{Du93,Park99,Morf02}.

In summary, we have shown that the quantum nature of the LLL crystal has striking implications for the defects. The lowest energy defect is a six-fold hyper-correlated bubble interstitial, the energy of which is a factor of three lower than the lowest defect in an uncorrelated Hartree-Fock crystal. The very low energy defects measured in experiments are thus a manifestation of the unusual quantum nature of the LLL crystal. 

We thank the National Science Foundation for support under Grant No. DMR-1005536 and the Research Computing and Cyberinfrastructure at Pennsylvania State University for providing high-performance computing resources. We thank Dr. Oleg Shklyaev for many excellent suggestions for the relaxation process, and acknowledge useful conversations with Profs. M. C. Cha and E. L. Altschuler.

%\bibliography{biblio_FQHE}

\begin{thebibliography}{37}
\expandafter\ifx\csname natexlab\endcsname\relax\def\natexlab#1{#1}\fi
\expandafter\ifx\csname bibnamefont\endcsname\relax
  \def\bibnamefont#1{#1}\fi
\expandafter\ifx\csname bibfnamefont\endcsname\relax
  \def\bibfnamefont#1{#1}\fi
\expandafter\ifx\csname citenamefont\endcsname\relax
  \def\citenamefont#1{#1}\fi
\expandafter\ifx\csname url\endcsname\relax
  \def\url#1{\texttt{#1}}\fi
\expandafter\ifx\csname urlprefix\endcsname\relax\def\urlprefix{URL }\fi
\providecommand{\bibinfo}[2]{#2}
\providecommand{\eprint}[2][]{\url{#2}}

\bibitem[{\citenamefont{Cross and Fisher}(1985)}]{Cross85}
\bibinfo{author}{\bibfnamefont{M.~C.} \bibnamefont{Cross}} \bibnamefont{and}
  \bibinfo{author}{\bibfnamefont{D.~S.} \bibnamefont{Fisher}},
  \bibinfo{journal}{Rev. Mod. Phys.} \textbf{\bibinfo{volume}{57}},
  \bibinfo{pages}{881} (\bibinfo{year}{1985}).

\bibitem[{\citenamefont{Leggett}(1970)}]{Leggett70}
\bibinfo{author}{\bibfnamefont{A.~J.} \bibnamefont{Leggett}},
  \bibinfo{journal}{Phys. Rev. Lett.} \textbf{\bibinfo{volume}{25}},
  \bibinfo{pages}{1543} (\bibinfo{year}{1970}).

\bibitem[{\citenamefont{Grimes and Adams}(1979)}]{Grimes79}
\bibinfo{author}{\bibfnamefont{C.~C.} \bibnamefont{Grimes}} \bibnamefont{and}
  \bibinfo{author}{\bibfnamefont{G.}~\bibnamefont{Adams}},
  \bibinfo{journal}{Phys. Rev. Lett.} \textbf{\bibinfo{volume}{42}},
  \bibinfo{pages}{795} (\bibinfo{year}{1979}).

\bibitem[{\citenamefont{Willett et~al.}(1988)\citenamefont{Willett, Stormer,
  Tsui, Pfeiffer, West, and Baldwin}}]{Willett88}
\bibinfo{author}{\bibfnamefont{R.~L.} \bibnamefont{Willett}},
  \bibinfo{author}{\bibfnamefont{H.~L.} \bibnamefont{Stormer}},
  \bibinfo{author}{\bibfnamefont{D.~C.} \bibnamefont{Tsui}},
  \bibinfo{author}{\bibfnamefont{L.~N.} \bibnamefont{Pfeiffer}},
  \bibinfo{author}{\bibfnamefont{K.~W.} \bibnamefont{West}}, \bibnamefont{and}
  \bibinfo{author}{\bibfnamefont{K.~W.} \bibnamefont{Baldwin}},
  \bibinfo{journal}{Phys. Rev. B} \textbf{\bibinfo{volume}{38}},
  \bibinfo{pages}{7881} (\bibinfo{year}{1988}).

\bibitem[{\citenamefont{Goldman et~al.}(1990)\citenamefont{Goldman, Santos,
  Shayegan, and Cunningham}}]{Goldman90}
\bibinfo{author}{\bibfnamefont{V.~J.} \bibnamefont{Goldman}},
  \bibinfo{author}{\bibfnamefont{M.}~\bibnamefont{Santos}},
  \bibinfo{author}{\bibfnamefont{M.}~\bibnamefont{Shayegan}}, \bibnamefont{and}
  \bibinfo{author}{\bibfnamefont{J.~E.} \bibnamefont{Cunningham}},
  \bibinfo{journal}{Phys. Rev. Lett.} \textbf{\bibinfo{volume}{65}},
  \bibinfo{pages}{2189} (\bibinfo{year}{1990}).

\bibitem[{\citenamefont{Jiang et~al.}(1990)\citenamefont{Jiang, Willett,
  Stormer, Tsui, Pfeiffer, and West}}]{Jiang90}
\bibinfo{author}{\bibfnamefont{H.~W.} \bibnamefont{Jiang}},
  \bibinfo{author}{\bibfnamefont{R.~L.} \bibnamefont{Willett}},
  \bibinfo{author}{\bibfnamefont{H.~L.} \bibnamefont{Stormer}},
  \bibinfo{author}{\bibfnamefont{D.~C.} \bibnamefont{Tsui}},
  \bibinfo{author}{\bibfnamefont{L.~N.} \bibnamefont{Pfeiffer}},
  \bibnamefont{and} \bibinfo{author}{\bibfnamefont{K.~W.} \bibnamefont{West}},
  \bibinfo{journal}{Phys. Rev. Lett.} \textbf{\bibinfo{volume}{65}},
  \bibinfo{pages}{633} (\bibinfo{year}{1990}).

\bibitem[{\citenamefont{Jiang et~al.}(1991)\citenamefont{Jiang, Stormer, Tsui,
  Pfeiffer, and West}}]{Jiang91}
\bibinfo{author}{\bibfnamefont{H.~W.} \bibnamefont{Jiang}},
  \bibinfo{author}{\bibfnamefont{H.~L.} \bibnamefont{Stormer}},
  \bibinfo{author}{\bibfnamefont{D.~C.} \bibnamefont{Tsui}},
  \bibinfo{author}{\bibfnamefont{L.~N.} \bibnamefont{Pfeiffer}},
  \bibnamefont{and} \bibinfo{author}{\bibfnamefont{K.~W.} \bibnamefont{West}},
  \bibinfo{journal}{Phys. Rev. B} \textbf{\bibinfo{volume}{44}},
  \bibinfo{pages}{8107} (\bibinfo{year}{1991}).

\bibitem[{\citenamefont{Paalanen et~al.}(1992)\citenamefont{Paalanen, Willett,
  Ruel, Littlewood, West, and Pfeiffer}}]{Paalanen92}
\bibinfo{author}{\bibfnamefont{M.~A.} \bibnamefont{Paalanen}},
  \bibinfo{author}{\bibfnamefont{R.~L.} \bibnamefont{Willett}},
  \bibinfo{author}{\bibfnamefont{R.~R.} \bibnamefont{Ruel}},
  \bibinfo{author}{\bibfnamefont{P.~B.} \bibnamefont{Littlewood}},
  \bibinfo{author}{\bibfnamefont{K.~W.} \bibnamefont{West}}, \bibnamefont{and}
  \bibinfo{author}{\bibfnamefont{L.~N.} \bibnamefont{Pfeiffer}},
  \bibinfo{journal}{Phys. Rev. B} \textbf{\bibinfo{volume}{45}},
  \bibinfo{pages}{13784} (\bibinfo{year}{1992}),
  \urlprefix\url{http://link.aps.org/doi/10.1103/PhysRevB.45.13784}.

\bibitem[{\citenamefont{Du et~al.}(1996)\citenamefont{Du, Tsui, Stormer,
  Pfeiffer, and West}}]{Du96}
\bibinfo{author}{\bibfnamefont{R.}~\bibnamefont{Du}},
  \bibinfo{author}{\bibfnamefont{D.}~\bibnamefont{Tsui}},
  \bibinfo{author}{\bibfnamefont{H.}~\bibnamefont{Stormer}},
  \bibinfo{author}{\bibfnamefont{L.}~\bibnamefont{Pfeiffer}}, \bibnamefont{and}
  \bibinfo{author}{\bibfnamefont{K.}~\bibnamefont{West}},
  \bibinfo{journal}{Solid State Commun.} \textbf{\bibinfo{volume}{99}},
  \bibinfo{pages}{755 } (\bibinfo{year}{1996}).

\bibitem[{\citenamefont{Pan et~al.}(2002)\citenamefont{Pan, Stormer, Tsui,
  Pfeiffer, Baldwin, and West}}]{Pan02}
\bibinfo{author}{\bibfnamefont{W.}~\bibnamefont{Pan}},
  \bibinfo{author}{\bibfnamefont{H.~L.} \bibnamefont{Stormer}},
  \bibinfo{author}{\bibfnamefont{D.~C.} \bibnamefont{Tsui}},
  \bibinfo{author}{\bibfnamefont{L.~N.} \bibnamefont{Pfeiffer}},
  \bibinfo{author}{\bibfnamefont{K.~W.} \bibnamefont{Baldwin}},
  \bibnamefont{and} \bibinfo{author}{\bibfnamefont{K.~W.} \bibnamefont{West}},
  \bibinfo{journal}{Phys. Rev. Lett.} \textbf{\bibinfo{volume}{88}},
  \bibinfo{pages}{176802} (\bibinfo{year}{2002}).

\bibitem[{\citenamefont{Santos et~al.}(1992)\citenamefont{Santos, Suen,
  Shayegan, Li, Engel, and Tsui}}]{Santos92}
\bibinfo{author}{\bibfnamefont{M.~B.} \bibnamefont{Santos}},
  \bibinfo{author}{\bibfnamefont{Y.~W.} \bibnamefont{Suen}},
  \bibinfo{author}{\bibfnamefont{M.}~\bibnamefont{Shayegan}},
  \bibinfo{author}{\bibfnamefont{Y.~P.} \bibnamefont{Li}},
  \bibinfo{author}{\bibfnamefont{L.~W.} \bibnamefont{Engel}}, \bibnamefont{and}
  \bibinfo{author}{\bibfnamefont{D.~C.} \bibnamefont{Tsui}},
  \bibinfo{journal}{Phys. Rev. Lett.} \textbf{\bibinfo{volume}{68}},
  \bibinfo{pages}{1188} (\bibinfo{year}{1992}).

\bibitem[{\citenamefont{Cs\'{a}thy et~al.}(2004)\citenamefont{Cs\'{a}thy, Tsui,
  Pfeiffer, and West}}]{Csathy04}
\bibinfo{author}{\bibfnamefont{G.~A.} \bibnamefont{Cs\'{a}thy}},
  \bibinfo{author}{\bibfnamefont{D.~C.} \bibnamefont{Tsui}},
  \bibinfo{author}{\bibfnamefont{L.~N.} \bibnamefont{Pfeiffer}},
  \bibnamefont{and} \bibinfo{author}{\bibfnamefont{K.~W.} \bibnamefont{West}},
  \bibinfo{journal}{Phys. Rev. Lett.} \textbf{\bibinfo{volume}{92}},
  \bibinfo{pages}{256804} (\bibinfo{year}{2004}).

\bibitem[{\citenamefont{Cs\'{a}thy et~al.}(2005)\citenamefont{Cs\'{a}thy, Noh,
  Tsui, Pfeiffer, and West}}]{Csathy05}
\bibinfo{author}{\bibfnamefont{G.~A.} \bibnamefont{Cs\'{a}thy}},
  \bibinfo{author}{\bibfnamefont{H.}~\bibnamefont{Noh}},
  \bibinfo{author}{\bibfnamefont{D.~C.} \bibnamefont{Tsui}},
  \bibinfo{author}{\bibfnamefont{L.~N.} \bibnamefont{Pfeiffer}},
  \bibnamefont{and} \bibinfo{author}{\bibfnamefont{K.~W.} \bibnamefont{West}},
  \bibinfo{journal}{Phys. Rev. Lett.} \textbf{\bibinfo{volume}{94}},
  \bibinfo{pages}{226802} (\bibinfo{year}{2005}).

\bibitem[{\citenamefont{Cs\'athy et~al.}(2007)\citenamefont{Cs\'athy, Tsui,
  Pfeiffer, and West}}]{Csathy07}
\bibinfo{author}{\bibfnamefont{G.~A.} \bibnamefont{Cs\'athy}},
  \bibinfo{author}{\bibfnamefont{D.~C.} \bibnamefont{Tsui}},
  \bibinfo{author}{\bibfnamefont{L.~N.} \bibnamefont{Pfeiffer}},
  \bibnamefont{and} \bibinfo{author}{\bibfnamefont{K.~W.} \bibnamefont{West}},
  \bibinfo{journal}{Phys. Rev. Lett.} \textbf{\bibinfo{volume}{98}},
  \bibinfo{pages}{066805} (\bibinfo{year}{2007}).

\bibitem[{\citenamefont{Yi and Fertig}(1998)}]{Yi98}
\bibinfo{author}{\bibfnamefont{H.}~\bibnamefont{Yi}} \bibnamefont{and}
  \bibinfo{author}{\bibfnamefont{H.~A.} \bibnamefont{Fertig}},
  \bibinfo{journal}{Phys. Rev. B} \textbf{\bibinfo{volume}{58}},
  \bibinfo{pages}{4019} (\bibinfo{year}{1998}).

\bibitem[{\citenamefont{Narevich et~al.}(2001)\citenamefont{Narevich, Murthy,
  and Fertig}}]{Narevich01}
\bibinfo{author}{\bibfnamefont{R.}~\bibnamefont{Narevich}},
  \bibinfo{author}{\bibfnamefont{G.}~\bibnamefont{Murthy}}, \bibnamefont{and}
  \bibinfo{author}{\bibfnamefont{H.~A.} \bibnamefont{Fertig}},
  \bibinfo{journal}{Phys. Rev. B} \textbf{\bibinfo{volume}{64}},
  \bibinfo{pages}{245326} (\bibinfo{year}{2001}).

\bibitem[{\citenamefont{Chang et~al.}(2005)\citenamefont{Chang, Jeon, and
  Jain}}]{Chang05}
\bibinfo{author}{\bibfnamefont{C.-C.} \bibnamefont{Chang}},
  \bibinfo{author}{\bibfnamefont{G.~S.} \bibnamefont{Jeon}}, \bibnamefont{and}
  \bibinfo{author}{\bibfnamefont{J.~K.} \bibnamefont{Jain}},
  \bibinfo{journal}{Phys. Rev. Lett.} \textbf{\bibinfo{volume}{94}},
  \bibinfo{pages}{016809} (\bibinfo{year}{2005}).

\bibitem[{\citenamefont{Archer and Jain}(2011)}]{Archer11}
\bibinfo{author}{\bibfnamefont{A.~C.} \bibnamefont{Archer}} \bibnamefont{and}
  \bibinfo{author}{\bibfnamefont{J.~K.} \bibnamefont{Jain}},
  \bibinfo{journal}{Phys. Rev. B} \textbf{\bibinfo{volume}{84}},
  \bibinfo{pages}{115139} (\bibinfo{year}{2011}).

\bibitem[{\citenamefont{Archer et~al.}(2013)\citenamefont{Archer, Park, and
  Jain}}]{Archer13}
\bibinfo{author}{\bibfnamefont{A.~C.} \bibnamefont{Archer}},
  \bibinfo{author}{\bibfnamefont{K.}~\bibnamefont{Park}}, \bibnamefont{and}
  \bibinfo{author}{\bibfnamefont{J.~K.} \bibnamefont{Jain}},
  \bibinfo{journal}{Phys. Rev. Lett.} \textbf{\bibinfo{volume}{111}},
  \bibinfo{pages}{146804} (\bibinfo{year}{2013}).

\bibitem[{\citenamefont{Fisher et~al.}(1979)\citenamefont{Fisher, Halperin, and
  Morf}}]{Fisher79}
\bibinfo{author}{\bibfnamefont{D.~S.} \bibnamefont{Fisher}},
  \bibinfo{author}{\bibfnamefont{B.~I.} \bibnamefont{Halperin}},
  \bibnamefont{and} \bibinfo{author}{\bibfnamefont{R.}~\bibnamefont{Morf}},
  \bibinfo{journal}{Phys. Rev. B} \textbf{\bibinfo{volume}{20}},
  \bibinfo{pages}{4692} (\bibinfo{year}{1979}).

\bibitem[{\citenamefont{Haldane}(1983)}]{Haldane83}
\bibinfo{author}{\bibfnamefont{F.~D.~M.} \bibnamefont{Haldane}},
  \bibinfo{journal}{Phys. Rev. Lett.} \textbf{\bibinfo{volume}{51}},
  \bibinfo{pages}{605} (\bibinfo{year}{1983}).

\bibitem[{\citenamefont{Thomson}(1904)}]{Thomson04}
\bibinfo{author}{\bibfnamefont{J.~J.} \bibnamefont{Thomson}},
  \bibinfo{journal}{Philos.Mag.} \textbf{\bibinfo{volume}{7}},
  \bibinfo{pages}{237} (\bibinfo{year}{1904}).

\bibitem[{tho()}]{thomprob06}
\bibinfo{note}{D.~J.~Wales and S.~Ulker, Phys. Rev. B {\bf 74}, 212101 (2006);
  D.~J.~ Wales, H.~McKay, and E.~L.~Altschuler, {\emph ibid.} {\bf 79}, 224115
  (2009). The minimum energy locations can be found at
  \url{http://thomson.phy.syr.edu/}}.

\bibitem[{\citenamefont{Koulakov et~al.}(1996)\citenamefont{Koulakov, Fogler,
  and Shklovskii}}]{Koulakov96}
\bibinfo{author}{\bibfnamefont{A.~A.} \bibnamefont{Koulakov}},
  \bibinfo{author}{\bibfnamefont{M.~M.} \bibnamefont{Fogler}},
  \bibnamefont{and} \bibinfo{author}{\bibfnamefont{B.~I.}
  \bibnamefont{Shklovskii}}, \bibinfo{journal}{Phys. Rev. Lett.}
  \textbf{\bibinfo{volume}{76}}, \bibinfo{pages}{499} (\bibinfo{year}{1996}).

\bibitem[{\citenamefont{Fogler et~al.}(1996)\citenamefont{Fogler, Koulakov, and
  Shklovskii}}]{Fogler96}
\bibinfo{author}{\bibfnamefont{M.~M.} \bibnamefont{Fogler}},
  \bibinfo{author}{\bibfnamefont{A.~A.} \bibnamefont{Koulakov}},
  \bibnamefont{and} \bibinfo{author}{\bibfnamefont{B.~I.}
  \bibnamefont{Shklovskii}}, \bibinfo{journal}{Phys. Rev. B}
  \textbf{\bibinfo{volume}{54}}, \bibinfo{pages}{1853} (\bibinfo{year}{1996}).

\bibitem[{\citenamefont{Price and Platzman}(1991)}]{Price91}
\bibinfo{author}{\bibfnamefont{R.}~\bibnamefont{Price}} \bibnamefont{and}
  \bibinfo{author}{\bibfnamefont{P.~M.} \bibnamefont{Platzman}},
  \bibinfo{journal}{Phys. Rev. B} \textbf{\bibinfo{volume}{44}},
  \bibinfo{pages}{2356} (\bibinfo{year}{1991}).

\bibitem[{\citenamefont{Cockayne and Elser}(1991)}]{Cockayne91}
\bibinfo{author}{\bibfnamefont{E.}~\bibnamefont{Cockayne}} \bibnamefont{and}
  \bibinfo{author}{\bibfnamefont{V.}~\bibnamefont{Elser}},
  \bibinfo{journal}{Phys. Rev. B} \textbf{\bibinfo{volume}{43}},
  \bibinfo{pages}{623} (\bibinfo{year}{1991}).

\bibitem[{\citenamefont{Maki and Zotos}(1983)}]{Maki83}
\bibinfo{author}{\bibfnamefont{K.}~\bibnamefont{Maki}} \bibnamefont{and}
  \bibinfo{author}{\bibfnamefont{X.}~\bibnamefont{Zotos}},
  \bibinfo{journal}{Phys. Rev. B} \textbf{\bibinfo{volume}{28}},
  \bibinfo{pages}{4349} (\bibinfo{year}{1983}).

\bibitem[{\citenamefont{Jain}(1989)}]{Jain89}
\bibinfo{author}{\bibfnamefont{J.~K.} \bibnamefont{Jain}},
  \bibinfo{journal}{Phys. Rev. Lett.} \textbf{\bibinfo{volume}{63}},
  \bibinfo{pages}{199} (\bibinfo{year}{1989}).

\bibitem[{\citenamefont{Jain}(2007)}]{Jain07}
\bibinfo{author}{\bibfnamefont{J.~K.} \bibnamefont{Jain}},
  \emph{\bibinfo{title}{Composite Fermions}} (\bibinfo{publisher}{Cambridge
  University Press, New York, US}, \bibinfo{year}{2007}).

\bibitem[{\citenamefont{Zheng and Fertig}(1994)}]{Zheng94}
\bibinfo{author}{\bibfnamefont{L.}~\bibnamefont{Zheng}} \bibnamefont{and}
  \bibinfo{author}{\bibfnamefont{H.~A.} \bibnamefont{Fertig}},
  \bibinfo{journal}{Phys. Rev. Lett.} \textbf{\bibinfo{volume}{73}},
  \bibinfo{pages}{878} (\bibinfo{year}{1994}).

\bibitem[{\citenamefont{Shi et~al.}(2008)\citenamefont{Shi, Jolad, Regnault,
  and Jain}}]{Shi08}
\bibinfo{author}{\bibfnamefont{C.}~\bibnamefont{Shi}},
  \bibinfo{author}{\bibfnamefont{S.}~\bibnamefont{Jolad}},
  \bibinfo{author}{\bibfnamefont{N.}~\bibnamefont{Regnault}}, \bibnamefont{and}
  \bibinfo{author}{\bibfnamefont{J.~K.} \bibnamefont{Jain}},
  \bibinfo{journal}{Phys. Rev. B} \textbf{\bibinfo{volume}{77}},
  \bibinfo{pages}{155127} (\bibinfo{year}{2008}).

\bibitem[{\citenamefont{Cha and Fertig}(1994{\natexlab{a}})}]{Cha94}
\bibinfo{author}{\bibfnamefont{M.-C.} \bibnamefont{Cha}} \bibnamefont{and}
  \bibinfo{author}{\bibfnamefont{H.~A.} \bibnamefont{Fertig}},
  \bibinfo{journal}{Phys. Rev. Lett.} \textbf{\bibinfo{volume}{73}},
  \bibinfo{pages}{870} (\bibinfo{year}{1994}{\natexlab{a}}).

\bibitem[{\citenamefont{Cha and Fertig}(1994{\natexlab{b}})}]{Cha94b}
\bibinfo{author}{\bibfnamefont{M.-C.} \bibnamefont{Cha}} \bibnamefont{and}
  \bibinfo{author}{\bibfnamefont{H.~A.} \bibnamefont{Fertig}},
  \bibinfo{journal}{Phys. Rev. B} \textbf{\bibinfo{volume}{50}},
  \bibinfo{pages}{14368} (\bibinfo{year}{1994}{\natexlab{b}}).

\bibitem[{\citenamefont{Du et~al.}(1993)\citenamefont{Du, Stormer, Tsui,
  Pfeiffer, and West}}]{Du93}
\bibinfo{author}{\bibfnamefont{R.~R.} \bibnamefont{Du}},
  \bibinfo{author}{\bibfnamefont{H.~L.} \bibnamefont{Stormer}},
  \bibinfo{author}{\bibfnamefont{D.~C.} \bibnamefont{Tsui}},
  \bibinfo{author}{\bibfnamefont{L.~N.} \bibnamefont{Pfeiffer}},
  \bibnamefont{and} \bibinfo{author}{\bibfnamefont{K.~W.} \bibnamefont{West}},
  \bibinfo{journal}{Phys. Rev. Lett.} \textbf{\bibinfo{volume}{70}},
  \bibinfo{pages}{2944} (\bibinfo{year}{1993}),
  \urlprefix\url{http://link.aps.org/doi/10.1103/PhysRevLett.70.2944}.

\bibitem[{\citenamefont{Park et~al.}(1999)\citenamefont{Park, Meskini, and
  Jain}}]{Park99}
\bibinfo{author}{\bibfnamefont{K.}~\bibnamefont{Park}},
  \bibinfo{author}{\bibfnamefont{N.}~\bibnamefont{Meskini}}, \bibnamefont{and}
  \bibinfo{author}{\bibfnamefont{J.}~\bibnamefont{Jain}}, \bibinfo{journal}{J.
  Phys. Condens. Mat.} \textbf{\bibinfo{volume}{11}} (\bibinfo{year}{1999}).

\bibitem[{\citenamefont{Morf et~al.}(2002)\citenamefont{Morf, d'Ambrumenil, and
  Das~Sarma}}]{Morf02}
\bibinfo{author}{\bibfnamefont{R.~H.} \bibnamefont{Morf}},
  \bibinfo{author}{\bibfnamefont{N.}~\bibnamefont{d'Ambrumenil}},
  \bibnamefont{and}
  \bibinfo{author}{\bibfnamefont{S.}~\bibnamefont{Das~Sarma}},
  \bibinfo{journal}{Phys. Rev. B} \textbf{\bibinfo{volume}{66}},
  \bibinfo{pages}{075408} (\bibinfo{year}{2002}),
  \urlprefix\url{http://link.aps.org/doi/10.1103/PhysRevB.66.075408}.

\end{thebibliography}

\end{document}